\documentclass{elsart}
\usepackage[english]{babel} 
\usepackage{times}
\usepackage[iso]{umlaute} 
\usepackage{graphicx}
\usepackage{paralist}
\usepackage{cite} 
\begin{document} 
\begin{frontmatter}

\title{Multiscale Modeling of Polymer Gels-Chemo-Electric Model versus Discrete Element Model} 

\author[1]{Thomas Wallmersperger\corauthref{cor}},
\corauth[cor]{Corresponding author. Tel.:+49-0711-685-7093 Fax.:+49-0711-685-3706}
 \ead{wittel@isd.uni-stuttgart.de}
\author[2]{Falk K. Wittel}
\author[1]{Michele D'Ottavio} and
\author[1]{Bernd-H. Kröplin}

\address[1]{Institute for Statics and Dynamics of Aerospace Structures, University of Stuttgart, Pfaffenwaldring 27, 70569 Stuttgart, Germany}
\address[2]{Institute for Building Materials (IfB), ETH Zurich, Switzerland} 

 \begin{abstract}
Polyelectrolyte gels are a very attractive class of actuation materials with remarkable electronic and mechanical properties with a great similarity to biological contractile tissues. They consist of a polymer network with ionizable groups and a liquid phase with mobile ions. Absorption and delivery of solvent lead to a large change of volume. This mechanism can be triggered by chemical (change of salt concentration or pH of solution surrounding the gel), electrical, thermal or optical stimuli. Due to this capability, these gels can be used as actuators for technical applications, where large swelling and shrinkage is desired.
In the present work chemically stimulated polymer gels in a solution bath are investigated. To adequately describe the different complicated phenomena occurring in these gels, they can be modeled on different scales. Therefore, models based on the statistical theory and porous media theory, as well as a coupled multi-field model and a discrete element formulation are derived and employed.

A refinement of the different theories from global macroscopic to microscopic are presented in this paper: The statistical theory is a macroscopic theory capable of describing the global swelling or bending, e.g., of a gel film, while the general theory of porous media (TPM) is a macroscopic continuum theory which is based on the theory of mixtures extended by the concept of volume fractions. The TPM is a homogenized model, i.e., all geometrical and physical quantities can be seen as statistical averages of the real quantities. The presented chemo-electro-mechanical multi-field formulation is a mesoscopic theory. It is capable of giving the concentrations and the electric potential in the whole domain. Finally the (micromechanical) discrete element (DE) theory is employed. In this case, the continuum is represented by distributed particles with local interaction relations combined with balance equations for the chemical field. This method is predestined for problems involving large displacements and strains as well as discontinuities.

In this paper, the coupled multi-field model and the discrete element model for chemical stimulation of a polymer gel film with and without domain deformation are employed. Based on these results, the presented formulations are compared and conclusions on their applicability in engineering practice are finally drawn.
\end{abstract}
\end{frontmatter}

\section{Introduction}
Polymer gels are ionic electroactive polymers. They consist of a polymer network with charged groups and a liquid phase with mobile ions, see Figure 1. By a change of the chemical milieu in the gel surrounding solution or by an applied electric field, the swelling of these gels can be triggered. Due to this capability, electrolyte polymer gels are designated as actuators for technical applications where large swelling and shrinkage is desired, such as for artificial muscles or other chemo-electro-mechanical actuators. The swelling behavior of these gels results from the equilibrium of different forces such as osmotic pressure forces, electrostatic forces and viscoelastic restoring forces.
\begin{figure}[htp] \centering{ \includegraphics[scale=0.85]{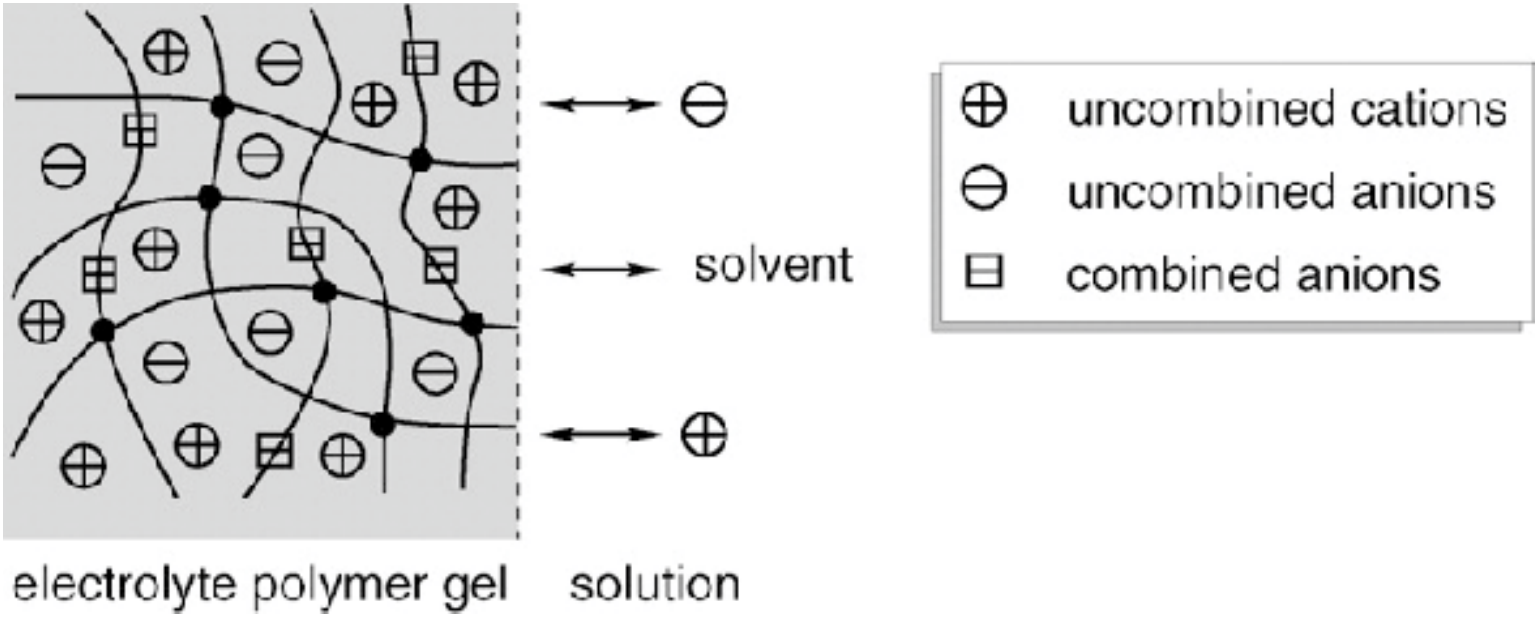} }
 \caption{Polymer gel network with bound anionic groups and mobile ions in solution} 
\end{figure}
\section{Modeling of Polymer Gels}
To describe the different phenomena adequately, the modeling is conducted on different scales. Starting on an atomistic level, the crystal structure of the material can be resolved; by increasing the length scale, the structure can be modeled by molecular dynamics or particle models which take into account the effects occurring on the microscale. By further increasing the scale, the whole structure can be investigated on the meso- or macroscale, e.g., when the prediction of the global failure of the structure is desired.

In the following, some key points on the statistical theory, the porous media theory, the coupled multi-field model and the discrete element theory are given.
\subsection{Statistical Theory}
The statistical theory is a method to describe the global swelling ratio of polyelectrolyte gels. This macroscopic theory is based on the swelling of network structures by Flory [1]. The conditions in the gel and of the gel surrounding solution is described by a change of the Gibbs free energy $\Delta F$. The total free energy is the sum of the free energies of mixing $\Delta F_M$ , of elastic deformation $\Delta F_{el}$ and of the mobile ion phases $\Delta F_{ion}$ ion inside and outside the gel, see [2, 3, 4]. The equilibrium state is characterized by a minimum of the total free energy difference $\Delta F$. In the standard form, this theory is valid for chemical and also thermal stimulation. More details on this theory together with numerical simulations can be found in the references given above or, e.g., in Wallmersperger et al. [5, 6] or Maurer et al. [7, 8].
\subsection{Porous Media Theory}
The general theory of porous media (TPM) is a macroscopic/mesoscopic continuum theory which is based on the theory of mixtures extended by the concept of volume fractions. In this theory neither the local porous micro structure nor the actual geometrical distribution of all the constituents have to be known [9, 10].

The TPM is formulated by conservation equations for the different phases or mixtures. It comprises local mechanical and chemical unknowns. The TPM is applicable first of all for the chemical stimulation but can be extended easily to other kinds of stimulation by adding new equations and coupling terms. More details can be found, e.g., in Ehlers et al. [11] and Kunz [12].
\subsection{Chemo-Electro-Mechanical Multi-Field Formulation}
The coupled multi-field formulation [13, 14] is formulated by different balance equations. It is capable of giving the local chemical, electrical and mechanical unknowns.

The chemical field is described by a convection(-migration)-diffusion equation for the different mobile species. The electric field is formulated by the Poisson equation. In contrast to the TPM, the electric field is directly obtained by solving the Poisson equation, i.e., no additional conditions prescribing the jump between gel and solution have to be used.

The mechanical field is formulated by the momentum equation. Normally, the deformation process is very slow, i.e., the influence of the inertia term is very small and the second order in time contributive term can thus be neglected.

The chemical and electrical field are solved simultaneously. The different concentrations in gel and solution give rise to an osmotic pressure term in the momentum equation leading to a change of the mechanical displacement.
\begin{figure}[htp] \centering{ \includegraphics[scale=0.85]{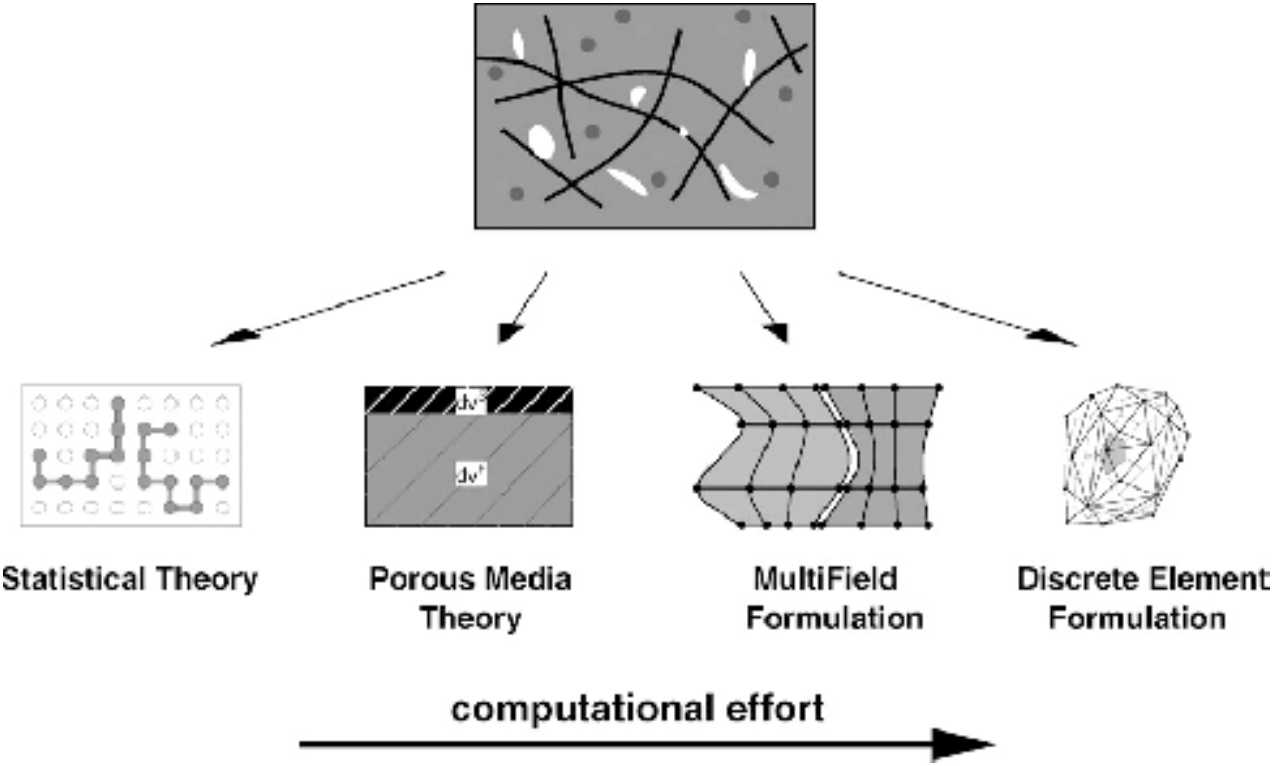} }
 \caption{Modeling strategies for electrolyte polymer gels} 
\end{figure}
\subsection{Discrete Element Theory}
The micromechanical behavior is described by the discrete element (DE) theory. In this model, the continuum is represented by distributed particles comprising a certain amount of mass. The particles interact with each other mechanically [15] by a superimposed truss or beam network of massless elements [16].

The mechanical behavior, i.e., the dynamics of the system, is obtained by solving the Newton's equations of motion
\begin{eqnarray}
m\ddot{x_i} &= \sum_j F_{ij} \\
\Theta \dot{\omega_i} &= \sum_j M_{ij}
\end{eqnarray}
where $\ddot{x}$ is the acceleration of the particles (in $x$-direction), $\dot{\omega}$ the angular acceleration, $m$ the mass of the particle, and $\Theta$ the rotational moment of inertia. $F_{ij}$ and $M_{ij}$ are the generalized forces and moments, respectively.

Additionally to the mechanical field, the chemical field, i.e., the ion movement inside the gel and from the gel to the solution, is described by diffusion equations for the different mobile particles. The concentration differences between the different regions of the gel and between gel and solution form the osmotic pressure difference, which is a main cause for the mechanical deformation of the polyelectrolyte gel film.
\section{Numerical Simulation}
In this section, the chemical stimulation, i.e., the change of salt concentration, of polymer gels in a solution bath is investigated.

In the first test case, the chemo-electrical behavior of a polyelectrolyte gel film (5x5 cm$^2$), fixed in a solution bath is considered (see Figure 3), while in the second test case the domain deformation is taken into account. In both 2D simulations, the concentration of the bound anionic groups in the gel is fixed. At $t = 0$ the concentration in the solution has been suddenly decreased from $c_{s,0} = 2$mM to $c_s = 1$mM. The parameters used in the numerical simulations are given in Table 1.
\begin{figure}[htp] \centering{ \includegraphics[scale=0.85]{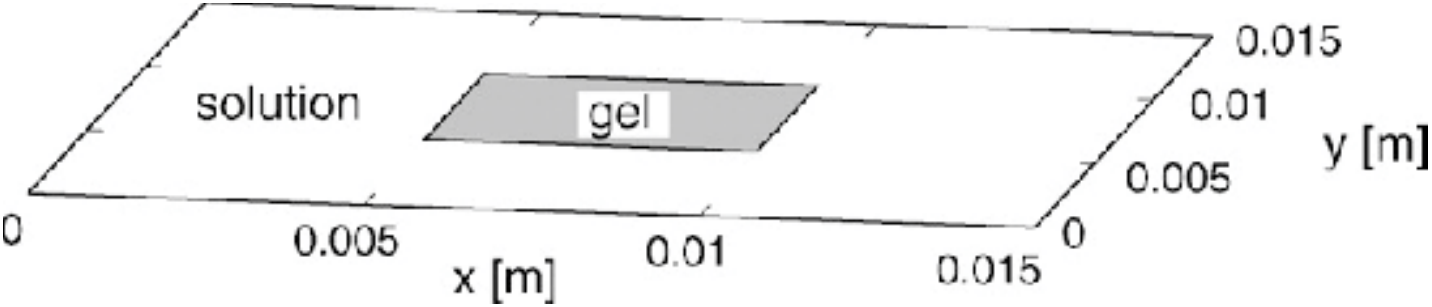} }
 \caption{Polymer gel in the solution bath} 
\end{figure}

\begin{table*}[htbp]
	\centering
		\begin{tabular}{p{2.5cm}p{2cm}|p{5cm}p{3.5cm}}
			Domain length: & $l_x$ = 15 mm	& Init. concentration of anionic rest groups: &	$c_{A_{,0}^-} = 2$mM \\
Domain width : &	$l_y$ = 15 mm &	Concentrations in the solution:	& $c_{Cl^-}^{(s)} = c_s = 1$mM \\
Gel length: &	$l_{xg}$ = 5 mm &	 &	$c_{Na+}^{(s)} = c_s = 1$mM \\
Gel width: &	$l_{yg}$ = 5 mm	& Initial concentrations in the solution: &	$c_{s,0} = 2$mM \\
Temperature: &	$T$ = 293 K &	Diffusion coefficients:	& $D = 10^{?7} $m$^2$/s \\
		\end{tabular}
	\caption{Parameters for the numerical simulation}
\end{table*}
\subsection{Chemo-Electrical Simulation on a Fixed Domain}
In this test case, only the chemo-electrical field will be investigated without considering any domain deformation.

The ion fluxes in the whole fixed gel-solution domain and the resulting time-history of the mobile concentrations and the potential difference between gel and solution are computed by using the coupled chemo-electric multi-field formulation.

With progressing time, the concentrations of $Na^+$ and $Cl^-$ in the gel are decreasing from the initial values ($c_{Na^+,0}^{(g)} = 3.2361$mM, $c_{Cl^{-}_{,0}}^{(g)} = 1.2361$mM) until the steady state ($c_{Na^+}^{(g)} = 2.4142$mM, $c_{Cl^{-}}^{(g)} =  1.4142$mM) is reached. In Figure 4 the stationary solution of the ion concentrations and the electric potential in the gel and the solution is depicted. The electroneutrality condition is fulfilled in both areas (gel and solution) outside the small boundary layer region.
\begin{figure}[htp] \centering{ \includegraphics[scale=0.85]{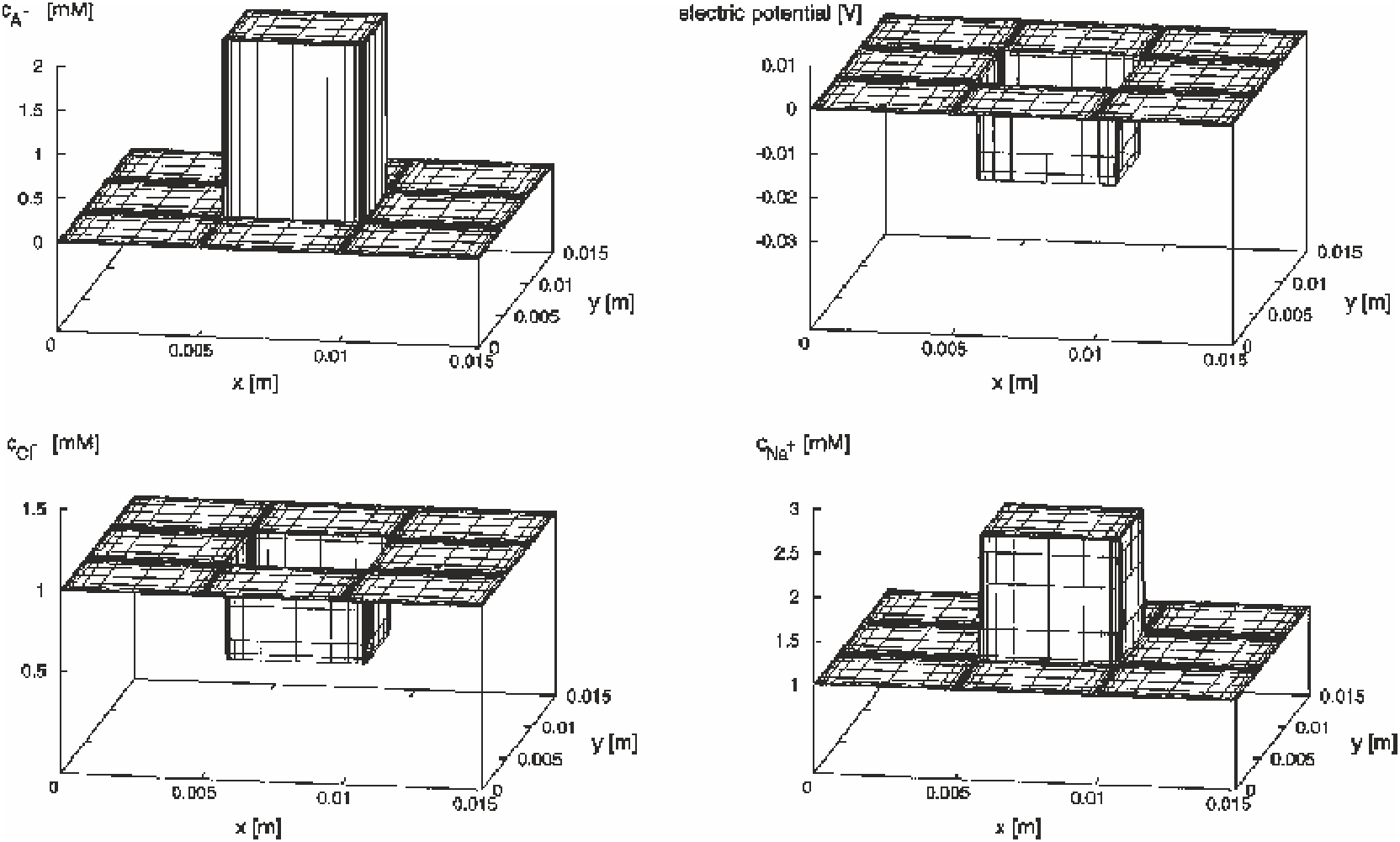} }
 \caption{2D-steady state results of the ion concentrations ($Cl^-$ (bottom left), $Na^+$ (bottom right)) and the electric potential $\psi$ in the gel and the solution without applied external electric field.} 
\end{figure}
\subsection{Discrete Element Simulation}
In the second test case, the swelling behavior of a polymer gel fiber in a solution bath is investigated with the discrete element (DE) method; i.e., the coupled chemo-electro-mechanical field is solved under consideration of the domain deformation.

For $t < 0$, the anionic gel ($c_{A_{,0}^-} = 2$mM) is located in a solution bath ($c_s = c_{Na^+}^{(s)} = c_{Cl^-}^{(s)} = 2$mM) in which the stationary equilibrium is fulfilled. At $t = 0$, the concentration in the solution bath is reduced to $c_s = 1$mM.

In the DE simulation, only the gel - i.e., without surrounding solution - is investigated. The boundary conditions at the gel/solution interface are computed at each time step by solving the neutrality condition in gel and solution
\begin{equation}
	\sum_\alpha(\zeta_\alpha C_\alpha)=0
\end{equation}
where $\zeta_\alpha$ is the charge of the ion species $\alpha$ - together with the Donnan-equilibrium
\begin{equation}
	\frac{c_{Na^+}^{(g)}}{c_{NA^+}^{(s)}}=\frac{c_{CL^-}^{(s)}}{c_{Cl^-}^{(g)}} \Leftrightarrow \frac{c_{Na^1}^{(g)}}{c_{s}}=\frac{c_{s}}{c_{Cl^-}^{(g)}} 
\end{equation}
resulting from identical electro-chemical potential in both regions. The resulting osmotic pressure $\Delta \pi$
\begin{equation}
	\Delta \pi RT\sum_{\alpha=1}^{N_f}(c_\alpha^{(s)}-c_\alpha^{(g)})
\end{equation}
due to the concentration differences leads to a change of the swelling ratio and therefore of the gel geometry. $R$ is the universal gas constant, $T$ the temperature and $c_\alpha^{(s)}$ and $c_\alpha^{(g)}$ are the concentrations of the species $\alpha$ in the solution and the gel, respectively.

Due to the fixed number of anionic charges, the number of moles in the gel remains identical, i.e., the concentration of bound groups varies by
\begin{equation}
	c_{A^-}=c_{A_{,0}^-}\frac{V_0}{V}
\end{equation}
where $V$ is the actual volume of the gel and $V_0$ the volume in reference state.

Since the boundary conditions are dependent on the salt concentration in the solution $c_s$ and the actual bound concentrations in the gel $c_{A^?}$, different values for the concentrations at the boundaries have to be prescribed for each time step.

In Figure 5, the initial conditions of the bound charges (top left) and of the mobile anions (top right) are depicted. Due to the reduction of the mobile ions in the solution, a new boundary condition at the gel/solution interface is prescribed and thus, the concentration of mobile anions in the gel is also decreasing. A typical plot of the mobile anions in the gel is given at a specific time $t_1$, see Figure 5 (middle right). Due to the resulting swelling of the gel film, the bound anionic concentration is reduced, see Figure 5 (middle left). The stationary solution of the bound anionic charges and mobile ions versus the new gel geometry is depicted in Figure 5 (bottom). Note, the cationic concentrations in the gel $c_{Na^+}^{(g)}$ may be calculated by
\begin{equation}
	c_{Na^+}^{(g)}=c_{Cl^-}^{(g)}+c_{A^-}
\end{equation}
\begin{figure}[htp] \centering{ \includegraphics[scale=0.75]{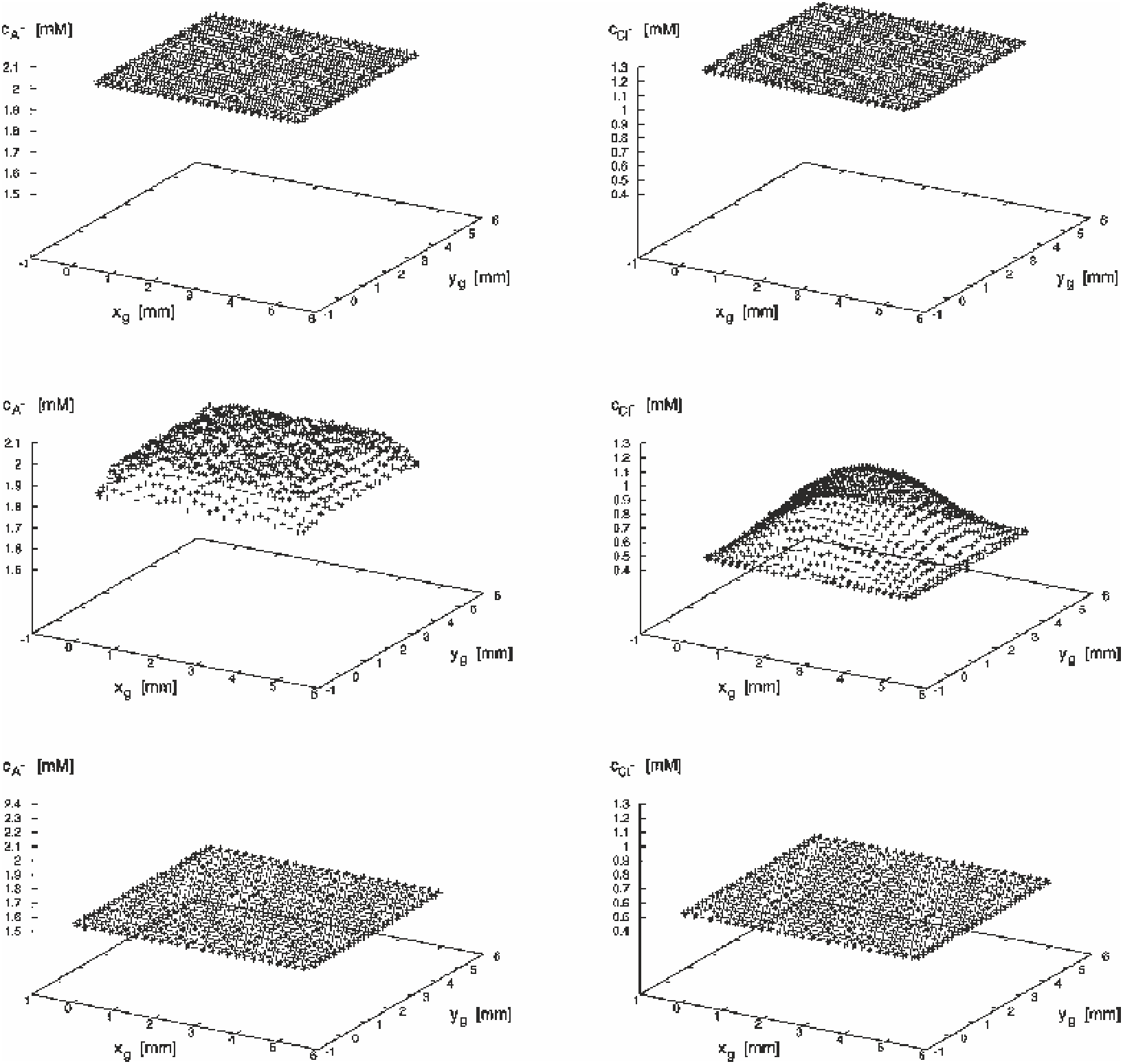} }
 \caption{Fixed negative charges (left) and mobile negative ions (right) versus the gel geometry at $t\leq 0$ (top), $t = t_1$ (middle) and $t\rightarrow\infty$ (bottom).} 
\end{figure}

The swelling of the gel is depicted in Figure 6: the initial geometry is represented by the black line while the swollen gel in steady-state is given by the grey one.
\begin{figure}[htp] \centering{ \includegraphics[scale=0.85]{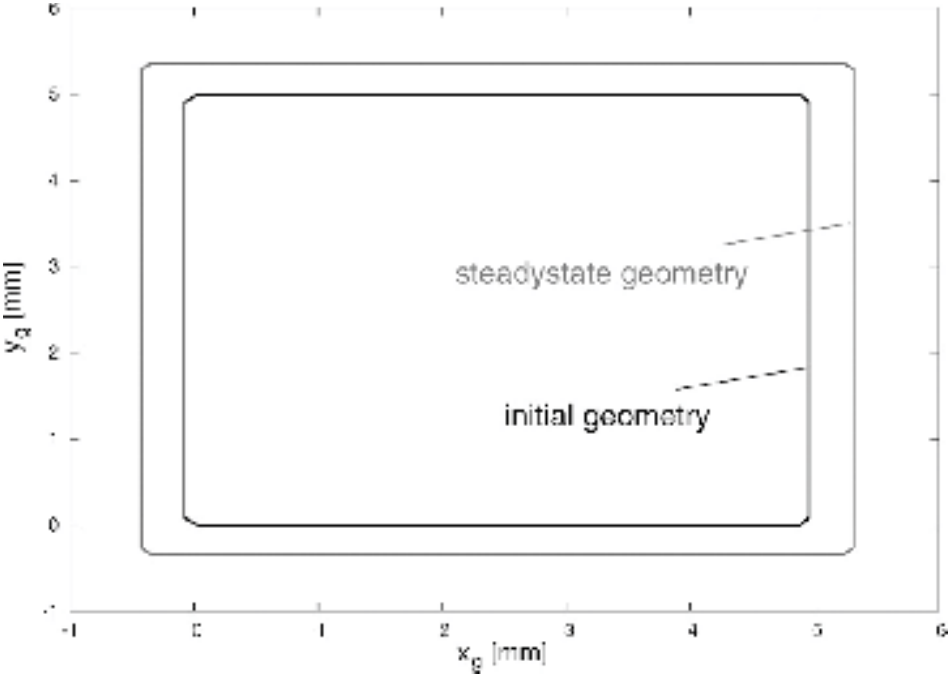} }
 \caption{Swelling of the gel fiber: gel geometry at $t <leq 0$ (black) and steady-state solution (grey).} 
\end{figure}
\subsection{Comparison}
In order to compare both test cases, in Figure 7, the concentrations of the fixed charges and the mobile ions in steady state without domain deformation (bottom left) and with domain deformation (bottom right) are depicted. The initial concentrations at $t < 0$ versus $x$ are given in Figure 7 (top).
\begin{figure}[htp] \centering{ \includegraphics[scale=0.85]{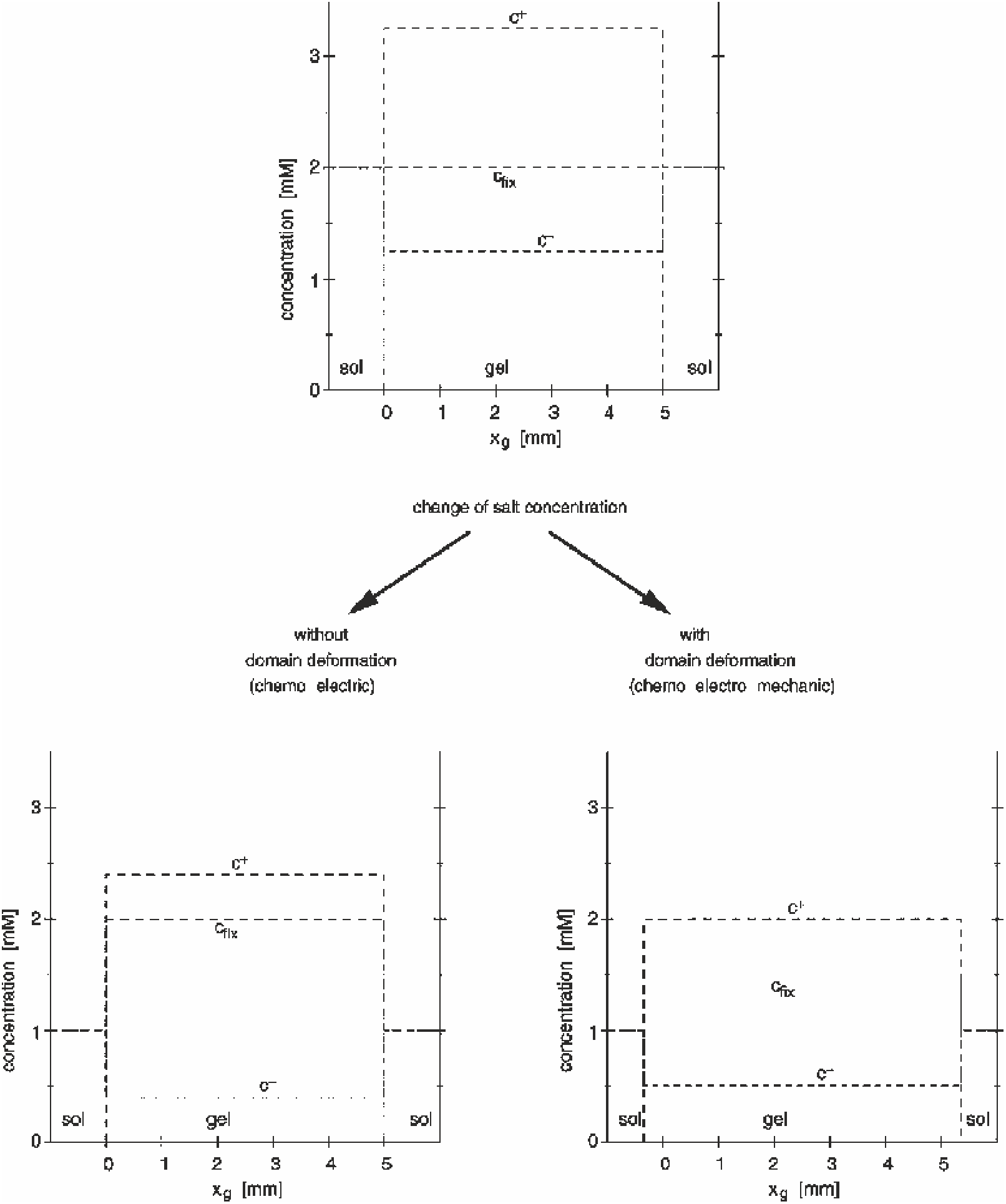} }F
 \caption{Concentrations of the bound anions $c_{fix} = c_{A^-}$ and the mobile cations $c^+ = c_{Na^+}$ and anions $c^- = c_{Cl^-}$ versus the $x_g$ -coordinate: initial state (top), steady state without swelling (bottom left) and steady state with swelling (bottom right).} 
\end{figure}

It can be seen that due to the resulting enlargement of the domain (Figure 7 (right)), the bound concentration $c_{A^-}$ decreases according to Eq. (6), while the total number of moles of bound charges remains constant. Due to the decrease of the bound anionic concentration, new concentration boundary conditions are prescribed - see Eqs. (4) and (7) - and therefore, the concentration of mobile cations in the gel also decreases. Summarizing, the positive domain deformation decreases the concentration differences; therefore the differential osmotic pressure decreases. This means, the enlargement of the gel domain is reduced by a kind of restoring force.
\section{Conclusion and Outlook}
Substantially different modeling strategies for the swelling behavior of polyelectrolyte gels under electro-chemical stimulation were presented in this paper.

If the global macroscopic, mechanic response is the only property of interest, the use of a simple statistical theory leads to sufficient results. Both the porous media theory and the multi-field formulation work within the assumptions and limitations of standard discretization schemes such as finite element or finite difference methods. Due to the field formulations, continuous fields of mechanical unknowns like displacements, strains and stresses, and of chemical and thermal unknowns are obtained in a straightforward way. These formulations are valid for quasi static and transient problems.

In order to model the chemical stimulation, for the statistical, the porous media as well as the discrete element theory, the gel is investigated only; therefore the Donnan equilibrium at the boundary has to be prescribed and the electro neutrality condition in the whole domain has to be fulfilled. The coupled multi-field formulation is applied for the whole gel-solution domain and the electroneutrality condition is not (directly) prescribed.

In this paper the coupled multi-field model and the discrete element model for chemical stimulation of a polymer gel film with and without domain deformation have been employed and compared. It can be seen that the chemo-electrical multi-field formulation is an optimal method to give the chemial and electrical unknowns in the whole domain. If the chemo-mechanical excitation should be considered, the discrete element approach in its present form is also predestined. The advantages of discrete element representations are, e.g., the direct physical access to the system during all the time steps of the simulation, and the representation of large deformations and strains.

\section*{References}
\begin{enumerate}[{[}1{]}]
\item Flory, P. J. 1953. Principles of Polymer Chemistry, Ithaca, NY: Cornell University Press.  
\item Ricka, J. and Tanaka, T. 1984. Swelling of ionic gels: Quantitative performance of the Donnan theory, Macromolecules, 17: 2917-2921.  
\item Ohmine, I. and Tanaka, T. 1982. Salt effects on the phase transition of ionic gels, Journal of Chemical Physics, 77(11): 5725-5729.  
\item Schröder, U. and Oppermann, W. 1996. Properties of polyelectrolyte gels, Physical Properties of Polymeric Gels, 19-38. John Wiley and Sons.  
\item Wallmersperger, T., Kröplin, B. and Gülch, R. W. 2004. Modelling and Analysis of Chemistry and Electromechanics, in Electroactive Polymer (EAP) Actuators as Artificial Muscles-Reality, Potential, and Challenges; Second Edition, 335-362. Bellingham, WA, USA: SPIE Press. Vol. PM 136 
\item Wallmersperger, T. 2003. Modellierung und Simulation stimulierbarer polyelektrolytischer Gele, PhD thesis Universität Stuttgart. Vol. 688 of Fortschritt-Berichte VDI: Reihe 5, Grund- und Werkstoffe, Kunststoffe, VDI-Verlag, 2003 
\item Maurer, G. and Prausnitz, J. M. 1996. Thermodynamics of phase equilibrium for systems containing gels, Fluid Phase Equilibria, 115: 113-133. 
\item Hüther, A., Xu, X. and Maurer, G. 2004. Swelling of n-isopropyl acrylamide hydrogels in water and aqueous solutions of ethanol and acetone, Fluid Phase Equilibria, 219: 231-244. 
\item Bowen, R. M. 1980. Incompressible porous media models by use of the theory of mixtures, Int. J. Engng. Sci., 18: 1129-1148. 
\item Ehlers, W. 2002. Foundations of multiphasic and porous materials, in Porous Media: Theory, Experiments and Numerical Applications, Edited by: Ehlers, W. and Bluhm, J. 3-86. Berlin: Springer-Verlag. 
\item Ehlers, W., Markert, B. and Acartürk, A., A continuum approach for 3-d finite viscoelastic swelling of charged tissues and gels, Proceedings of Fifth World Congress on Computational Mechanics. Edited by: Mang, H. A., Rammerstorfer, F. G. and Eberhardsteiner, J. 
\item Kunz, W. 2003. Mehrpasenmodell zur Beschreibung ionischer Gele im Rahmen der Theorie poröser Medien, Master's thesis Universität Stuttgart.
\item Wallmersperger, T., Kröplin, B., Holdenried, J. and Gülch, R. W., A coupled multi-field formulation for ionic polymer gels in electric fields, 8th International Symposium on Smart Structures and Materials, Vol. 4329 of Electroactive Polymer Actuators and Devices. Edited by: Bar-Cohen, Y. pp.264-275. SPIE.
\item Wallmersperger, T., Kröplin, B. W. and Gülch, R. 2004. Coupled chemo-electro-mechanical formulation for ionic polymer gels-numerical and experimental investigations, Mechanics of Materials, 36(5-6): 411-412. 
\item Johnson, K. L. 2001. Contact Mechanics, Cambridge Univ. Press.
\item Hrennikoff, A. 1941. Solution of problems of elasticity by the framework method, Journal of Applied Mechanics, 8(4): A169-A175.
\end{enumerate}
\end{document}